\documentclass[12pt]{article}
\usepackage{bbm}
\usepackage{epsfig}
\usepackage{array}
\usepackage{float}
\usepackage{dsfont}
\usepackage{amstext}
\usepackage{cite}

\usepackage{a4}

\usepackage{a4wide}

\def\ra{\rightarrow}
\def\be{\begin{equation}}
\def\ee{\end{equation}}
\def\gs{\mathrel{
   \rlap{\raise 0.511ex \hbox{$>$}}{\lower 0.511ex \hbox{$\sim$}}}}
\def\ls{\mathrel{
   \rlap{\raise 0.511ex \hbox{$<$}}{\lower 0.511ex \hbox{$\sim$}}}}
\newcommand{\obb}{0\mbox{$\nu\beta\beta$}}
\newcommand{\onbb}{neutrino-less double beta decay}
\newcommand{\ba}{\begin{array}{c}}
\newcommand{\baz}{\begin{array}{cc}}
\newcommand{\bad}{\begin{array}{ccc}}
\newcommand{\bav}{\begin{array}{cccc}}
\newcommand{\baf}{\begin{array}{ccccc}}
\newcommand{\bea}{\begin{equation} \begin{array}{c}}
\newcommand{\eea}{ \end{array} \end{equation}}
\newcommand{\ea}{\end{array}}
\newcommand{\D}{\displaystyle}
\newcommand{\dms}{\mbox{$\Delta m^2_{\odot}$}}
\newcommand{\dma}{\mbox{$\Delta m^2_{\rm A}$}}
\newcommand{\meff}{\mbox{$\langle m \rangle$}}
\newcommand{\imeff}{\mbox{$\D \left\langle \frac{1}{m} \right\rangle$}}

\begin{document}

\title{
\hfill {\small arXiv: 0912.3388 [hep-ph]}
\vskip .4cm
\bf \Large 
On Non-Unitary Lepton Mixing and Neutrino Mass Observables}
\author{\\  
Werner Rodejohann\thanks{email: 
\tt werner.rodejohann@mpi-hd.mpg.de}
\\ \\
{\normalsize \it Max--Planck--Institut f\"ur Kernphysik,}\\
{\normalsize \it  Postfach 103980, D--69029 Heidelberg, Germany} 
}

\date{}
\maketitle
\begin{abstract}
\noindent  
There are three observables related to neutrino mass, namely 
the kinematic mass in direct searches, the effective 
mass in neutrino-less double beta decay, and the sum of 
neutrino masses in cosmology. In the limit of exactly degenerate neutrinos 
there are very simple relations between those observables, and we 
calculate corrections due to non-zero mass splitting. 
We discuss how the possible non-unitarity of the lepton 
mixing matrix may modify these relations 
and find in particular that corrections due to non-unitarity can exceed 
the corrections due to mass splitting. 
We furthermore investigate 
constraints from neutrino-less double beta decay on  
mass and mixing parameters of heavy neutrinos in the 
type I see-saw mechanism. There are constraints  
from assuming that heavy neutrinos are exchanged, and constraints 
from assuming light neutrino exchange, which 
arise from an exact see-saw relation. The latter has its origin in 
the unitarity violation arising in see-saw scenarios. 
We illustrate that 
the limits from the latter approach are much stronger. The drastic 
impact on inverse neutrino-less double 
beta decay $(e^- e^- \ra W^- W^-)$ is studied. 
We furthermore discuss neutrino mixing in case there is  
one or more light sterile neutrino.  
Neutrino oscillation probabilities 
for long baseline neutrino oscillation experiments are considered,  
and the analogy to general non-unitarity phenomenology, 
such as zero-distance effects, is pointed out.

\end{abstract}

\newpage

\section{\label{sec:intro}Introduction}
Even though neutrino mass and lepton mixing are firmly established
facts, there is a plethora of open issues still to be addressed. 
Two of those questions concern the scale of neutrino mass and whether 
the lepton mixing matrix is unitary. In this letter we would like to discuss 
in particular some aspects that arise when one combines these two aspects.\\ 

The surprisingly large amount of phenomenology of a non-unitarity 
lepton mixing matrix has recently been discussed by 
various authors
\cite{old0,old,SA,CP_Sp,GO,AM,Xing_meff,WR,xing,others}. 
The observable consequences range from non-standard phenomena in 
neutrino oscillation experiments \cite{CP_Sp,GO,AM} 
to modified properties of leptogenesis \cite{WR}. One particular
aspect, which so far has only been partly discussed in 
Refs.~\cite{Xing_meff,xing},
is the impact of non-unitarity on observables related to neutrino mass. 
There are three experimental avenues and measurable
quantities to pin down neutrino mass, 
namely\footnote{The precise definition of these quantities will be given
in Section \ref{sec:mass}.}  (i) direct searches 
probing $m_\beta$; (ii) \onbb~probing \meff; (iii) cosmology probing 
$\Sigma$. In the testable case of quasi-degenerate neutrinos with a mass scale 
$m_0$, and when one neglects the small mass splittings, unitarity of
the lepton mixing matrix leads to the relation 
$\meff^{\rm max} = m_\beta = \Sigma/3$, where  $\meff^{\rm max}$
denotes the maximal value of \meff. We discuss how non-unitarity of
the mixing matrix affects this relation and compare these effects with
the inevitable effect of non-zero mass splitting. Surprisingly, the
effect of non-unitarity can lead to larger corrections, 
though the overall effect is small of course.  

We furthermore revisit the content of Ref.~\cite{Xing_meff}, which 
uses present limits from \onbb~(\obb) to test mass and mixing parameters of 
heavy neutrinos in the type I see-saw mechanism. 
The limits which are obtained
when one assumes that light neutrinos are exchanged in the 
\obb-diagram can (in the type I see-saw) 
be exactly related to heavy neutrino parameters. 
This is possible because the lepton mixing matrix in see-saw scenarios 
is non-unitary. 
One can compare these limits with the constraints
arising when one assumes that these heavy neutrinos are exchanged in
the \obb-diagram. We illustrate that the limits on the heavy 
mass and mixing parameters 
from light neutrino exchange are much stronger than the ones from 
heavy neutrino exchange. The impact of 
this approach on inverse neutrino-less double 
beta decay $(e^- e^- \ra W^- W^-)$ is also studied, and it is shown 
that the process is heavily suppressed.  

Finally, we discuss the presence of light sterile neutrinos in
addition to the three active ones. In this situation 
the $3\times3$ matrix which describes the mixing of active 
neutrinos amongst each other is 
in general not unitary. We show how to identify the mixing parameters 
of the sterile neutrinos with the small parameters commonly used to
parameterize a non-unitary mixing matrix $N$, namely in $N = (1 + \zeta) \,
U_0$, where $U_0$ is unitary and $\zeta$ hermitian. We illustrate how
these parameters enter the observables related to neutrino mass. 
Moreover, we note the analogy of this realization of non-unitarity, 
which is kinematically easily accessible, to the usually considered one, 
for which it is assumed that the reason for non-unitarity is beyond 
kinematic reach. 
For instance, in case of eV$^2$ scale mass-squared 
differences corresponding to the sterile neutrinos, 
``zero-distance effects'' in 
neutrino oscillation probabilities arise due to the averaged-out 
oscillations associated with these mass-squared differences. \\

The letter is build up as follows: in Section \ref{sec:mass} we
discuss observables related to neutrino mass and their
connection in case the mixing matrix is unitary. In Section
\ref{sec:NU} we let non-unitary enter the game and discuss how this
influences the relations among the mass observables. The particular example
of type I see-saw parameters and \obb~is 
discussed there as well. Finally, 
in Section \ref{sec:NUs} the analogy between light sterile neutrinos 
and non-unitarity is treated, before we conclude in 
Section \ref{sec:concl}.

\section{\label{sec:mass}Neutrino Mass Observables}

In the charged lepton basis the 
Pontecorvo-Maki-Nakagawa-Sakata (PMNS) matrix $U$ diagonalizes 
the neutrino mass matrix $m_\nu$ and is 
parameterized as 
\be \label{eq:Upara} 
U = 
\left( 
\bad  
c_{12}  c_{13} & s_{12}  c_{13} & s_{13}
e^{- i \delta}
\\[0.2cm] 
-s_{12}  c_{23} - c_{12}  s_{23} 
 s_{13}   e^{i \delta} 
& c_{12}   c_{23} - s_{12}   s_{23}  
 s_{13}  e^{i \delta} 
& s_{23}   c_{13} 
\\[0.2cm] 
s_{12}   s_{23} - c_{12}   c_{23}  
 s_{13}  
e^{i \delta} & 
- c_{12}  s_{23} - s_{12}   c_{23}  
 s_{13}   e^{i \delta} 
& c_{23}  c_{13} 
\ea 
\right) P \, .
\ee 
\noindent Here 
$c_{ij} = \cos \theta_{ij}$, $s_{ij} = \sin \theta_{ij}$ 
and the Majorana phases are contained in the diagonal matrix 
$P = {\rm diag}(1, \, e^{i\alpha_2/2} , \, e^{i\alpha_3/2})$. 
The eigenvalues of $m_\nu$ are the neutrino masses and 
there are three complementary observables related to them. 
The most model-independent one is the kinematic mass as measurable in
beta decay experiments: 
\be \label{eq:mb}
m_\beta = \sqrt{\sum |U_{ei}|^2 \, m_i^2 } \, .
\ee 
The current limit is 2.3 eV \cite{mainz} and with the 
KATRIN experiment a 
90 \% C.L.~limit of $m_\beta \le 0.2$ eV will be possible, while a discovery
potential of $5\sigma$ for $m_\beta = 0.35$ eV exists \cite{katrin}. 
Presumably unfair to other experiments, we will denote the observable 
$m_\beta$ in the following as  
the KATRIN observable. 

In contrast to the incoherent sum in Eq.~(\ref{eq:mb}), 
neutrino-less double beta decay (\obb) is sensitive 
to the coherent sum (the ``effective mass'') 
\be \label{eq:meff}
\meff = \left| \sum U_{ei}^2 \, m_i \right|  .
\ee
The measured limits on the half-life of \onbb~are 
around $10^{23}$ and $10^{25}$~y, depending on 
the nucleus \cite{APS}. Depending also on the uncertain 
nuclear matrix elements, 
upper limits on \meff~around 0.5 to 1 eV arise if one assumes 
that light neutrino exchange is responsible for \obb. 
The existing limits on the half-lifes will be 
improved considerably (by two orders of magnitude or more) 
in the near future by various experiments \cite{APS}. 
The maximal value of the effective mass,  
\be
\meff^{\rm max} = \sum \left|U_{ei}\right|^2 \, m_i  \, , 
\ee
 is obtained when all Majorana phases are trivial (e.g., zero) 
and the three now real terms in the sum of Eq.~(\ref{eq:meff}) 
simply add. 

Finally, the sum of neutrino masses,  
\be \label{eq:sigma}
\Sigma \equiv \sum m_i \, , 
\ee
can be extracted from cosmological observations and is bounded from
above by about 1 eV, depending on the data set, priors and of course the
cosmological model \cite{cosmo}. One expects future limits on 
$\Sigma$ in the 0.1 eV range. 

We can easily relate the different observables in the limit 
of quasi-degenerate neutrinos. 
If all mass splittings are set to zero, and the common neutrino
mass is denoted with $m_0$, then the relation 
\be \label{eq:naive}
m_0 = m_\beta = \meff^{\rm max} = \Sigma/3 
\ee
is immediately obtained when unitarity of $U$ is assumed\footnote{There 
are similar relations with the
minimal value of \meff, which however depend on the neutrino mixing
parameters $|U_{e1}|$ and $|U_{e3}|$. Since these parameters are presently 
not known exactly (see e.g.~\cite{bari}) we do not consider these
relations. We also do not consider normal or inversely 
hierarchical mass scenarios, in which the effect we will study is less 
emphasized, but experimentally even more difficult to probe.}. 
Mismatch of these relations would for quasi-degenerate neutrinos 
indicate that some form of new physics is present, 
see e.g., \cite{MMR}. It is known that 
there are corrections to relations (\ref{eq:naive}) when 
non-zero mass splitting is
taken into account \cite{split}. 

We will evaluate these corrections in the following: 
our convention is such that $m_0$ is always the heaviest neutrino mass. 
Hence, in the normal ordering we denote $m_0 = m_3$, while for the
inverted ordering $m_0 = m_2$. Defining the
quantities 
\be \label{eq:eta}
\eta_\odot = \frac{\dms}{2 \, m_0^2} ~\mbox{ and }~
\eta_{\rm A} = \frac{\dma}{2 \, m_0^2} \, , 
\ee
we can obtain the following expressions for the masses 
\be 
\bav 
\mbox{normal:} & m_3 = m_0 ~,
& m_2 \simeq m_0 \left(1 - \eta_{\rm A} - \frac 12  \,
\eta_{\rm A}^2 \right) ~,
& m_1 \simeq m_0 \left(1 - \eta_{\rm A} - \frac 12  \,
\eta_{\rm A}^2 - \eta_\odot \right) \, ,\\ 
\mbox{inverted:} & m_2 = m_0 ~,
& m_1 \simeq m_0 \left(1 -   \eta_\odot\right) ~,& 
m_3 \simeq m_0 \left(1 + \eta_{\rm A} - \frac 12 \, 
\eta_{\rm A}^2 - \eta_\odot \right) .
\ea
\ee
We have kept here second order terms for the ratio $\eta_{\rm A}$ 
of the atmospheric 
mass-squared difference and $m_0^2$, which can be of the same order of
magnitude as the terms $\eta_\odot$ of order solar mass-squared 
difference divided by $m_0^2$. 
This form of expansion will be applied for most of this work, and we 
mention here already that we will later neglect also terms of order 
$\eta_{\rm A}^2 \,|U_{e3}|^2$, because they correspond 
in magnitude to terms of order $\eta_\odot^2$.    
With this expansion, the results for the three observables are 
(we start with assuming a normal mass ordering)  
\bea \label{eq:NO} \D 
m_\beta \simeq m_0 \left(1 -  c_{13}^2 \, \eta_{\rm A}- \frac 12 \,  
c_{13}^4 \, \eta_{\rm A}^2 - c_{12}^2 \, c_{13}^2 \, \eta_\odot 
 \right) \equiv \tilde{m}_\beta \, , \\ \D 
\meff^{\rm max} \simeq m_\beta , \\ \D 
\Sigma \simeq 3 \, m_0 \left( 
1 - \frac 23 \, \eta_{\rm A} - \frac{1}{3}\, \eta_{\rm A}^2 - \frac 13 \, 
 \eta_\odot 
\right) \equiv \tilde{\Sigma} \, . 
\eea
We see that the upper limit of the effective mass and the KATRIN 
observable are to the order given identical. Their difference is by all means 
tiny, to be more precise: 
\be \label{eq:NO_diff} \D 
 m_\beta - \meff^{\rm max} \simeq  m_0 \left( 
\frac 14 \, \sin^2 2 \theta_{12} \,  
\eta_\odot^2 +  |U_{e3}|^2 \, c_{12}^2 \, \eta_\odot \, 
\eta_{\rm A} + \frac 12 \, 
|U_{e3}|^2 \, \eta_{\rm A}^2 \right) \equiv m_0 \, \Delta_1 
\, . 
\ee
Numerically, if we choose $m_0 = 0.3$ eV and insert \cite{bari} 
$\dma = 2.39 \cdot 10^{-3}$ eV$^2$, $\dms = 7.67 \cdot 10^{-5}$ eV$^2$ 
and $\sin^2 \theta_{12} = \frac 13$, then $\Delta_1$ is 
$2.1 \cdot 10^{-8} $ for $U_{e3} = 0$ and $3.6 \cdot 10^{-6}$ 
for $|U_{e3}| = 0.2$. Any deviation from these numbers will therefore 
be a clear signal of new physics.

The situation is somewhat different when we consider $\Sigma$. 
In the fully degenerate regime, we
have the relation $\Sigma = 3 \, m_\beta$. Corrections are induced by 
non-zero splitting, which results in  
\be
\frac 13 \, \Sigma - m_\beta \simeq 
\frac 13 \, \Sigma - \meff^{\rm max} \simeq \frac{m_0}{6} \left( 
(3 \, \cos 2 \theta_{13} - 1) \, \eta_{\rm A} + 
\eta_{\rm A}^2 + (1 + 3 \, \cos 2 \theta_{12}) 
\, \eta_\odot \right) \equiv m_0 \, \Delta_2 \, . 
\ee
With the same numerical input as above for $\Delta_1$, the range of 
$\Delta_2$ is between 0.0046 for $|U_{e3}| = 0$ and 0.0041 for 
$|U_{e3}| = 0.2$. 

In the inverted ordering, the formulae are basically identical and 
in particular the size of $m_\beta - \meff^{\rm max}$, as well as of 
$\frac 13 \, \Sigma - m_\beta$ is identical to the normal ordering, except for 
a sign change for the latter.

\section{\label{sec:NU}Non-Unitarity and Neutrino Mass Observables}

\subsection{\label{sec:NUl}General Case}
Now let us switch on non-unitarity, or rather switch off unitarity.  
It proves convenient to 
write in the relation $\nu_\alpha = N_{\alpha i} \, \nu_i$, 
which connects flavor and mass states, the now non-unitary matrix $N$ 
as \cite{CP_Sp}
\be \label{eq:Ueta}
N = (1 + \eta) \, U_0\,,
\ee   
where $\eta$ is hermitian (containing 6 real moduli 
and 3 phases because $\eta_{\alpha \beta} = |\eta_{\alpha \beta}| 
\, e^{i \phi_{\alpha \beta}}$ for $\alpha \neq \beta$) 
and $U_0$ is unitary (containing 3 real moduli 
and 3 phases). 
Several observables \cite{SA} such as limits on rare lepton decays and 
universality 
lead to the following 90\% C.L.~bounds on the elements of 
$\eta$:  
\be
|\eta| = 
\left( \bad 
|\eta_{ee}| & |\eta_{e\mu}| & |\eta_{e\tau}| \\
\cdot & |\eta_{\mu\mu}| & |\eta_{\mu\mu}| \\
\cdot & \cdot & |\eta_{\mu\tau}|
\ea
\right) \le 
\left( \bad 
2.0 \times 10^{-3} & 5.9 \times 10^{-5} & 1.6 \times 10^{-3} \\
\cdot & 8.2 \times 10^{-3} & 1.0 \times 10^{-3} \\
\cdot & \cdot & 2.6 \times 10^{-3}
\ea
\right).
\ee 
These bounds apply to sources of non-unitarity corresponding to
energies much larger than experimentally accessible. Mixing of light 
neutrinos with heavy (say, with masses $\gs$ TeV) 
particles is one example for such a situation. 
Using now $N$ instead of $U$, we can re-evaluate the results from 
the last Section, in particular the difference of 
the KATRIN observable $m_\beta$ and the maximal value of the 
effective mass for a normal mass ordering. 
We find that the parameter $\eta_{ee}$ related to non-unitarity 
contributes in first order\footnote{
In a specific parametrization for unitarity violation in the type I 
see-saw mechanism, formulas for \meff~and $m_\beta$ have been given in 
\cite{xing}.}: 
\bea \label{eq:NO_NU}
m_\beta \simeq \tilde{m}_\beta + m_0 \, |\eta_{ee}| \, \\
\meff^{\rm max} \simeq \tilde{m}_\beta + 2 \, m_0 \, |\eta_{ee}| \, .  
\eea
The term $\tilde{m}_\beta$ has been defined above 
in Eq.~(\ref{eq:NO}).  
The terms $\eta_{e\mu}$ and $\eta_{e \tau}$ contribute only quadratically, 
the other entries of $\eta$ do not show up at all. 
Hence, the difference between the effective mass and the KATRIN observable 
is 
\be 
m_\beta - \meff^{\rm max} \simeq m_0 \, ( \Delta_1 
 - |\eta_{ee}| )\, ,
\ee 
where $\Delta_1$ is defined in Eq.~(\ref{eq:NO_diff}). 
Applying our example of $m_0 = 0.3$ eV, we find that the term 
$m_0 \, |\eta_{ee}|$ can be as large as 
$6.0 \cdot 10^{-4}$ eV, which has to be compared with 
$m_0 \, \Delta_1$ being mostly of order $10^{-6}$ eV.  
Though the correction from $|\eta_{ee}|$ is 
below current experimental sensitivities, it is surprising 
and noteworthy that it can be almost 3 orders of magnitude larger than the 
correction from mass splitting.

The sum of masses does not depend on the mixing matrix 
and is therefore not 
affected by its non-unitarity. 
For the differences of $\Sigma$ 
with the KATRIN observable and the maximal effective mass one finds 
\bea \D 
\frac{1}{3} \, \Sigma - m_\beta \simeq m_0 \, (\Delta_2 - |\eta_{ee}|) \, ,\\ \D
\frac{1}{3} \, \Sigma - \meff^{\rm max} \simeq m_0 \, (\Delta_2 - 2 \, 
|\eta_{ee}|) \, . 
\eea
With our example values discussed above $\Delta_2$ is smaller than 
0.0046 and thus we see (after recalling that $|\eta_{ee}| < 0.0020$)  
that also the relations $\Sigma/3 = m_\beta$ and 
$\Sigma/3= \meff^{\rm max}$ can be modified 
by non-unitarity to a larger extent than due to mass-splitting. 

In the inverted ordering one finds the same corrections to the differences 
of observables than for the normal ordering.\\

\subsection{\label{sec:NUh}Heavy Neutrino Mixing in the 
type I See-Saw}

Within the conventional type I see-saw there is a guaranteed 
source of unitarity
violation, namely the inherent mixing of the light neutrino states
with the heavy ones. The complete mass term containing the Dirac and Majorana 
masses can be written as 
\be
{\cal L} = \overline{\nu_L} \, m_D \, N_R + \frac 12 \, 
\overline{N_R^c} \, M_R \, N_R + h.c. 
= 
\frac 12 \, (\overline{\nu_L}, \overline{N_R^c}) \left( 
\baz  
0 & m_D \\
m_D^T & M_R
\ea \right) 
\left(\ba \nu_L^c \\ N_R \ea \right) + h.c.,    
\ee 
and it is diagonalized by a unitary $6\times6$ matrix 
\be \label{eq:Uss}
{\cal U} = 
\left( \baz 
N & S \\
T & V 
\ea \right) \mbox{ with } 
{\cal M} = 
 {\cal U} 
 \left( 
\baz 
m_\nu^{\rm diag} & 0 \\
0 & M_R^{\rm diag} 
\ea \right) {\cal U}^T \, .
\ee
Since the eigenvalues of $M_R$ are much bigger than the elements of
$m_D$, the entries of $S$ and $T$ are of order $m_D/M_R$, and hence 
one can obtain the expression 
\be 
-N^\dagger \, m_D \, M_{R}^{-1} \, m_D^T \, N^\ast 
= m_\nu^{\rm diag} \, .
\ee
The matrix $S$ characterizes the mixing of the light 
neutrinos with the heavy ones:
\be 
\nu_\alpha = N_{\alpha i} \, \nu_i + S_{\alpha i} \, N_i \, ,
\ee
where $\nu_i$ ($N_i$) are the light (heavy) neutrinos with 
$i = 1,2,3$ and $\alpha = e, \mu, \tau$. 
Therefore, the mixing matrix in type I see-saw scenarios 
is strictly speaking not unitary, since 
$N N^\dagger = \mathbbm{1} - S S^\dagger\neq \mathbbm{1}$. 

The effective mass in neutrino-less double beta decay is given by 
\be
\meff = \left| \sum N_{ei}^2 \, m_i \right| \le \xi~{\rm eV}\, , 
\ee
and has, as discussed above, 
a current limit of around 1 eV, depending on the uncertain 
nuclear matrix elements. We have written therefore the limit as 
$\xi$ eV, with $\xi = {\cal O}(1)$. 

To connect the effective mass with the heavy neutrino parameters, 
one notes that the 11-element of Eq.~(\ref{eq:Uss}) reads 
$N \, m_\nu^{\rm diag} \, N^T + S \, M_R^{\rm diag} \, S^T = 0$. 
Therefore, for the effective mass holds \cite{Xing_meff} 
\be
\left|\sum N_{ei}^2 \, m_i\right|  
= \left|\sum S_{ei}^2 \, M_i\right| \, . 
\ee 
Consequently, the experimental 
limits on \meff~apply directly to this combination
of parameters: 
\be \label{eq:xing}
\left|\sum S_{ei}^2 \, M_i \right| \le \xi ~\rm eV \, . 
\ee 
This relation has been discussed first in Ref.~\cite{Xing_meff}, and here 
we illustrate its phenomenological consequences further.  
Note that this relation has its origin in the necessary unitary violation 
of the PMNS matrix in type I see-saw scenarios. 
The combination of parameters in Eq.~(\ref{eq:xing}) has to be
compared with another combination of mass and mixing parameters, 
which can be constrained from \obb~\cite{imeff}: 
\be \label{eq:weak}
\imeff = \left| \sum S_{ei}^2 \, \frac{1}{M_i} \right| 
\le \tilde \xi ~ 5 \cdot 10^{-8} \, {\rm GeV}^{-1} \, . 
\ee
This limit (we have introduced a factor $\tilde \xi$ taking into account
possible nuclear matrix element uncertainties) 
is obtained when \obb~is assumed to be mediated by heavy
neutrinos\footnote{For an analysis of similar limits 
in other processes, see \cite{imeffs}.}. 
The origin of the difference between 
\meff~and \imeff~is nothing but the two extreme limits of 
the fermion propagator of the Majorana neutrinos, which is 
central to the Feynman diagram of \obb: 
\be
\frac{{\slash \hspace{-.2cm}q} + m_i}{q^2 - m_i^2} 
\propto \left\{ 
\baz 
m_i & \mbox{ for } q^2 \gg m_i^2 \\
\frac{\D 1}{\D m_i} & \mbox{ for } q^2 \ll m_i^2
\ea
\right. \, .
\ee
Here $q$ denotes the momentum transfer in the process under study, 
which is around $10$ to $100$ MeV, corresponding to $1/r$, where 
$r \simeq 10^{-12}$ cm is the average distance of the 
two decaying nuclei\footnote{We note that typically the contribution 
from light neutrino exchange is larger by at least five 
orders of magnitude than the 
one from heavy neutrino exchange \cite{Hax,Xing_meff}. Hence, in order 
to apply the limit on \imeff~one has to assume that 
$\meff$~is very close to zero, i.e., a normal hierarchy 
must necessarily be present.}. 

\begin{figure}[t]
\begin{center}
\includegraphics[width=14cm,height=9cm]{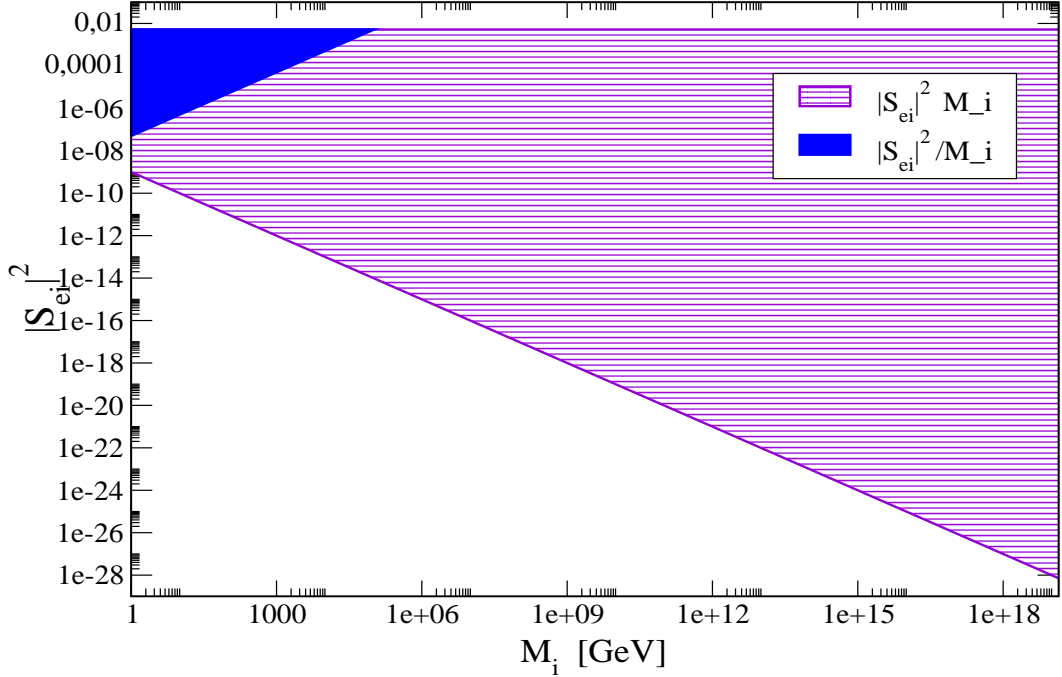} 
\caption{\label{fig:limit} Constraints on heavy Majorana neutrino
parameters from neutrino-less double beta decay. 
The horizontally striped (violet) area is forbidden by  
$|S_{ei}|^2 \, M_i < 1 $ eV, while the filled (blue) 
area is forbidden by $|S_{ei}|^2 /M_i < 5 \cdot 10^{-8}$ GeV$^{-1}$. 
The general upper 
limit is $|S_{ei}|^2 < 0.0052$.}
\end{center}
\end{figure}

We see that $\meff$ and $\imeff$ depend on two different 
combinations of the mass and mixing parameters of the heavy 
neutrinos. It is a useful exercise 
to compare the two approaches. Fig.~\ref{fig:limit} shows the 
two limits. In what regards the mixing parameter $S_{ei}$, there 
is an upper 
limit of \cite{nardi}
\be
|S_{ei}|^2 \le 0.0052 \, ,
\ee
obtained from global fits, in particular of LEP data. 
The horizontally striped (violet) 
area in Fig.~\ref{fig:limit} is obtained from 
$|S_{ei}|^2 \, M_i < 1 $ eV, while the filled (blue) 
area is from $|S_{ei}|^2 /M_i < 5 \cdot 10^{-8}$ GeV$^{-1}$. 
It is clear to see that the constraints from 
$|S_{ei}|^2 \, M_i < 1 $ eV are much stronger.

 We should note that the limits from $|S_{ei}|^2 \, M_i$ apply to heavy 
neutrinos of the see-saw mechanism, while limits from 
$|S_{ei}|^2/M_i$ apply to any heavy neutral Majorana fermion which mixes  
with the light fermions.  
We have also assumed that there is only one heavy neutrino 
contributing to the 
sum, and therefore do not take the possibility of fine-tuned 
cancellations of different terms in the sum into account.
We also do not take into account fine-tuned cancellations in 
neutrino-less double beta decay, i.e., there could 
be diagrams (e.g.~right-handed currents, supersymmetry,$\ldots$) 
contributing to \obb, 
which interfere destructively with the neutrino mass contribution(s).\\

\begin{figure}[th]
\begin{center}
\includegraphics[width=8cm,height=7cm]{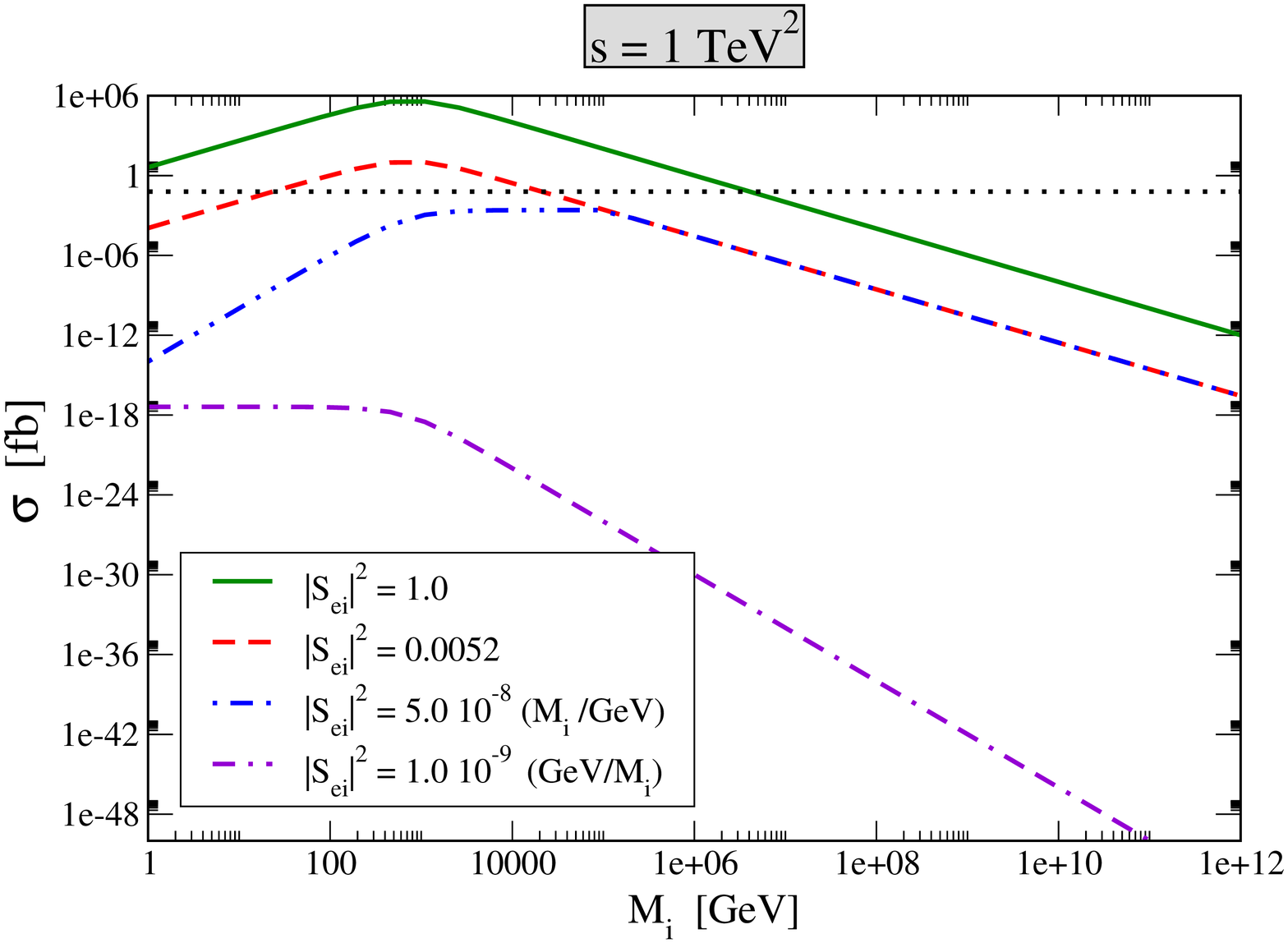} 
\includegraphics[width=8cm,height=7cm]{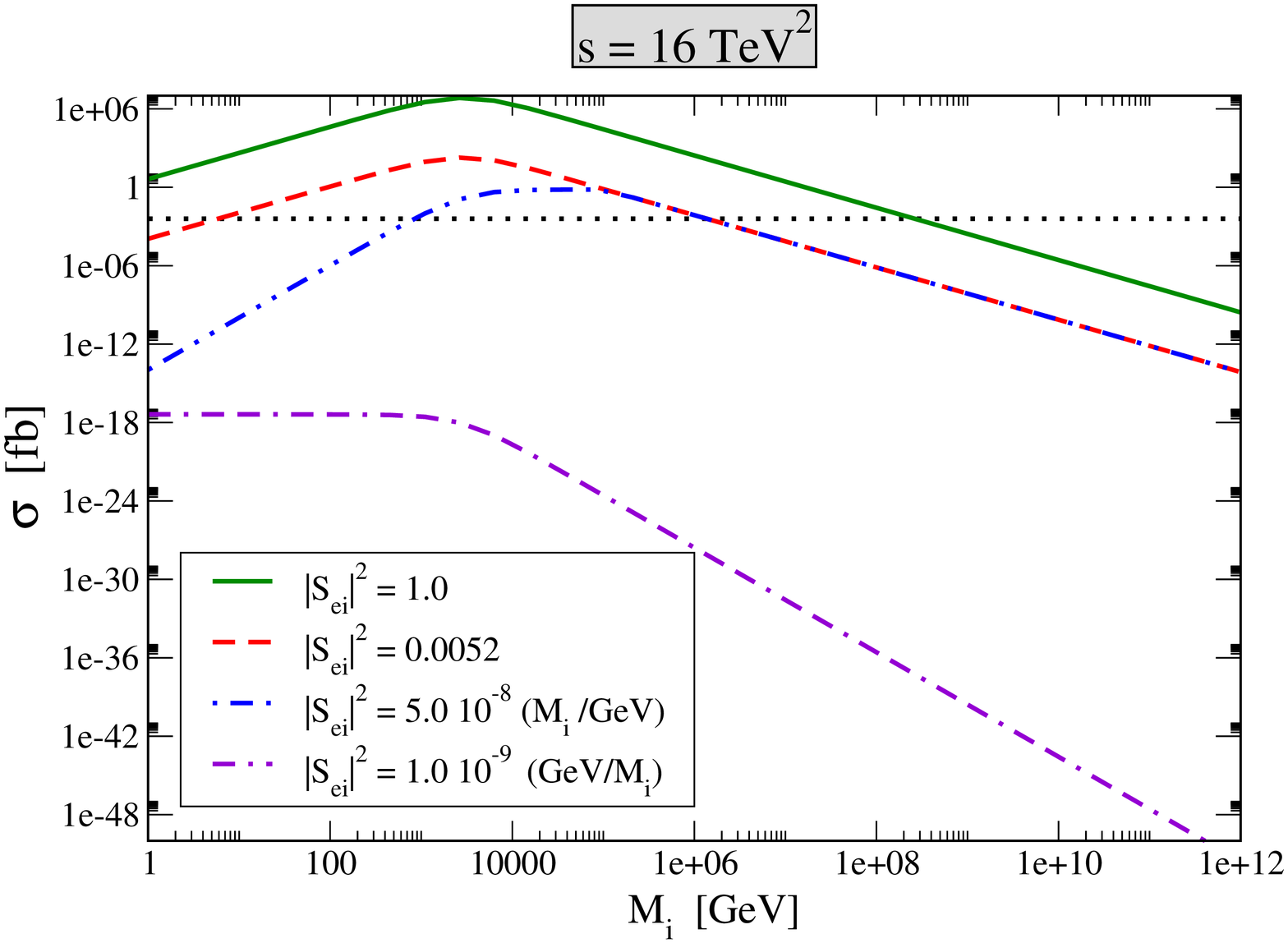}
\includegraphics[width=8cm,height=7cm]{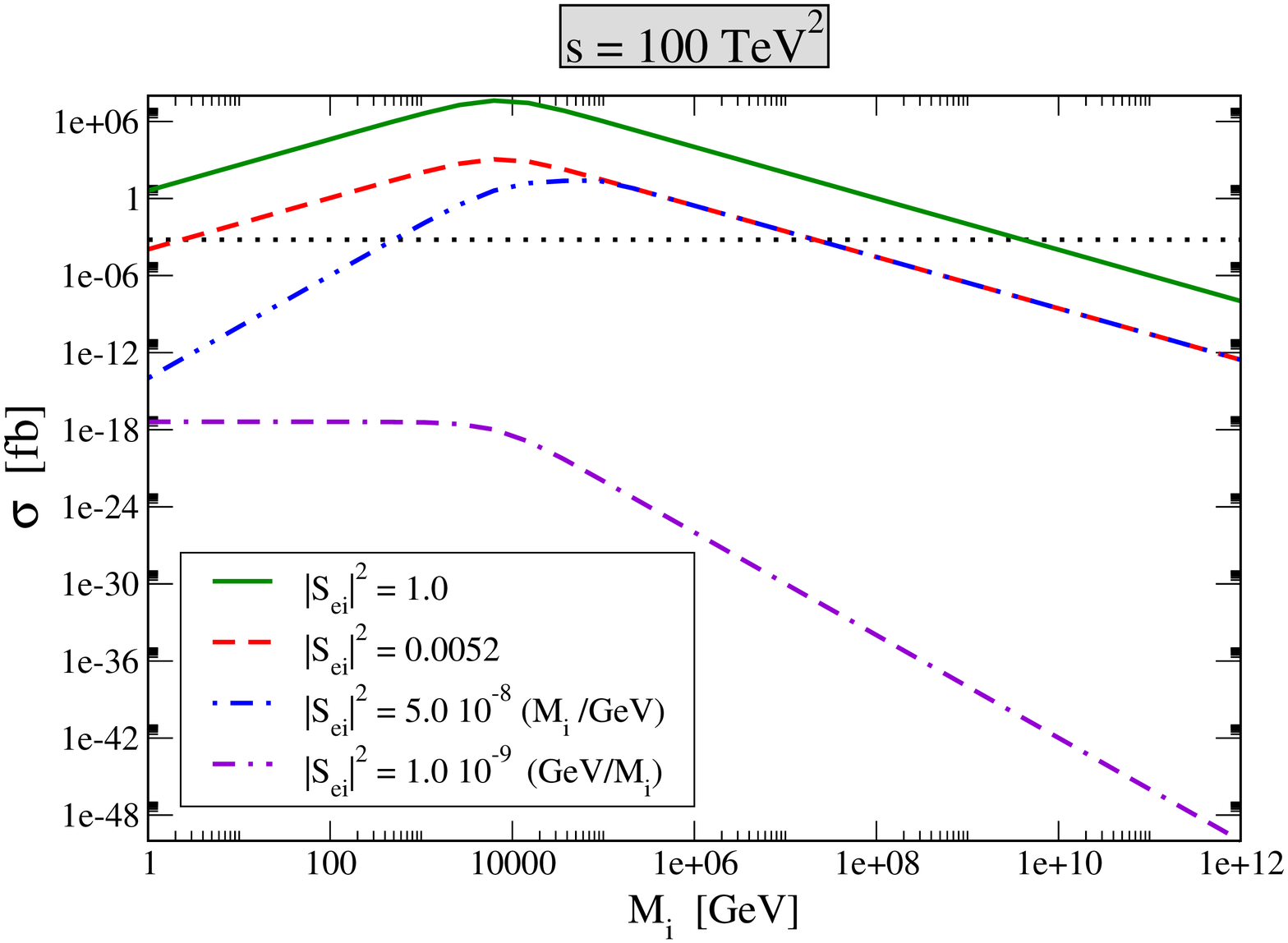}
\caption{\label{fig:2}Cross section for $e^- e^- \ra W^- W^-$ with 
$\sqrt{s} = 1$ TeV (upper left), $\sqrt{s} = 4$ TeV (upper right), 
$\sqrt{s} = 10$ TeV (bottom) and several limits for the mixing parameter 
$|S_{ei}|^2$. The dotted line corresponds to five events for an 
assumed luminosity of 80 $(s/{\rm TeV^2})$ fb$^{-1}$.}
\end{center}
\end{figure}

With these caveats in mind, we continue with studying a frequently 
considered process related to collider phenomenology of heavy
Majorana neutrinos. There are many possibilities for this, but here 
we focus for simplicity on ``inverse neutrino-less 
double beta decay'' \cite{i0,belanger},   
\[ 
e^- e^- \ra W^- W^- \, ,
\] 
which may be searched for at future linear colliders in an 
electron-electron mode. The differential cross section in the limit 
of negligible mass of the $W^-$ and one heavy neutrino $M_i$ is 
\be
\frac{d \sigma}{d \cos \theta} = 
\frac{G_F^2}{32 \, \pi } \left\{ M_i \, |S_{ei}|^2 \left( 
\frac{t}{t - M_i^2} + \frac{u}{u - M_i^2} 
\right) \right\}^2 \, .
\ee
Evaluating the full cross section for $\sqrt{s} = 1$ TeV, 
 4 TeV, and 10 TeV results in 
Fig.~\ref{fig:2}, where we give $\sigma$ for different values of 
$|S_{ei}|^2$. We have also indicated where to expect 
five events when a luminosity of 80 $(s/{\rm TeV^2})$ 
fb$^{-1}$ \cite{belanger} is achieved. Note that the extreme limits of the
cross section, which are given as 
\be
\sigma = \left\{ 
\baz \D 
\frac{G_F^2}{4 \, \pi} \, \left(|S_{ei}|^2 \, M_i\right)^2 & \mbox{for } 
s \gg M_i^2 \, , \\ \D 
\frac{G_F^2}{16 \, \pi}  s^2 
\left(\frac{|S_{ei}|^2}{M_i}\right)^2 & \mbox{for } 
s \ll M_i^2 \, ,
\ea \right. 
\ee
do not depend on $s$ for negligible masses and are proportional to 
$s^2$ for heavy masses.

The effect of the strong limit $|S_{ei}|^2 \, M_i < 1 $
eV is again clearly seen. In case of $\sqrt{s} = 1$ TeV even the 
weaker limit (\ref{eq:weak}) 
from \obb~renders the process unobservable. 
For the larger values of $\sqrt{s} = 4$ and 10 TeV a large range of 
values of $M_i$ could potentially be probed. However, switching on 
the limit (\ref{eq:xing}) from the see-saw relation makes the 
cross section miniscule. This straightforward example shows that the 
limit (\ref{eq:xing}) makes it much harder to observe any 
effect of Majorana neutrinos at colliders (if they are see-saw based and 
no cancellations take place).

\section{\label{sec:NUs}Light sterile Neutrinos and Non-Unitarity}

Another straightforward source of non-unitarity of the $3\times3$ 
lepton mixing matrix is the presence of light sterile neutrinos, 
with masses in the eV regime.  
We wish to remark in this note a nice analogy between the effects of 
sterile neutrinos and of kinematically unaccessible non-unitarity, which 
takes place when neutrino oscillations are considered. 

Assuming first the presence of only one sterile neutrino, 
the total light neutrino mass matrix is now a $4\times4$ object, 
and diagonalized by a unitary $4\times4$ matrix 
\be \label{eq:Ut}
U = \left( 
\bav 
\multicolumn{3}{c}{\mbox{$N_{3 \times 3}$}} & U_{e4} \\
U_{s1} & U_{s2} & U_{s3} & U_{s4}
\ea 
\right) . 
\ee
The ``active'' $3\times3$ part $N$, which describes the mixing of 
$\nu_e$, $\nu_\mu$ and $\nu_\tau$, is not unitarity, because 
$N N^\dagger \neq  \mathbbm{1}_3$. We parameterize the full 
$4\times4$ mixing matrix as 
\be \label{eq:Ust}
U = \tilde{R}_{14} \, R_{24} \, \tilde{R}_{34} \, 
R_{23} \, \tilde{R}_{13} \, R_{12} \, P \, ,
\ee
where $P = {\rm diag}(1,e^{i\alpha_2/2},
e^{i\alpha_3/2},e^{i\alpha_4/2})$ contains the Majorana phases and 
$R_{ij}$ are rotations around the $ij$-axis, e.g., 
\be
R_{23} = 
\left( 
\bav 
1 & 0 & 0 & 0 \\
0 & c_{23} & s_{23} & 0 \\
0 & -s_{23} & c_{23} & 0 \\
0 & 0 & 0 & 1
\ea 
\right),~~
\tilde{R}_{14}
= 
\left( 
\bav 
c_{14} & 0 & 0 & s_{14} \, e^{-i \delta_{14}} \\
0 & 1 & 0 & 0 \\
0 & 0 & 1 & 0 \\
-s_{14} \, e^{i \delta_{14}} & 0 & 0 & 1
\ea 
\right).
\ee
The upper left $3\times3$ submatrix of $U$ is called $N$. 
Recall that one can write $N = (1 + \zeta) \, U_0$ with 
unitary $U_0$ and hermitian $\zeta$, and hence we 
can identify $2 \, \zeta \simeq N  N^\dagger - \mathbbm{1}$. 
Expanding in terms of the small mixing angles $\theta_{i4}$, 
we find from Eq.~(\ref{eq:Ust}): 
\be
\zeta \simeq -\frac 12 
\left( 
\bad 
\theta_{14}^2 & \theta_{14} \, \theta_{24} \, e^{-i \delta_{14}} 
& \theta_{14} \, \theta_{34} \, e^{i (\delta_{34} -\delta_{14})} \\
\theta_{14} \, \theta_{24} \, e^{i \delta_{14}} & 
\theta_{24}^2  & \theta_{24} \, \theta_{34} \, e^{i \delta_{34}} \\
\theta_{14} \, \theta_{34} \, e^{-i (\delta_{34} -\delta_{14})} & 
\theta_{24} \, \theta_{34} \, e^{-i \delta_{34}} & 
\theta_{34}^2 
\ea
\right) .
\ee
We should stress here that the limits on $\eta$ quoted in 
Eq.~(\ref{eq:eta}) do of course not apply to the elements of $\zeta$. 
The limits on $\eta$ are valid for particles which are
kinematically far from being accessible. 
We note that if the LSND result \cite{lsnd} is analyzed in 
terms of neutrino oscillations, the angles $\theta_{i4}$ 
are of order $0.1$. However, in terms of oscillation 
experiments there are 
similar effects from sterile neutrinos and 
from a non-unitary mixing matrix, as we will note in the following.

To illustrate this, consider that the sterile neutrinos have typically  
mass-squared differences of order eV$^2$, i.e., much larger than the 
solar and atmospheric mass-squared difference. It suffices to 
make our point for the simple case of a high-energy, short-baseline 
neutrino factory with baseline $L = 130$ km (CERN--Frejus) and a muon 
energy of $E_\mu = 50$ GeV. One can neglect matter effects and  
the solar neutrino mass-squared difference, 
hence set $\dms = m_2^2 - m_1^2$ to zero, while the oscillations 
corresponding to $\Delta m^2_{41} \simeq 
\Delta m^2_{42} \simeq \Delta m^2_{43} \simeq $ eV$^2$ 
average out. The oscillation probability is in general 
\be 
P_{\alpha \beta} = \left| 
\sum_j U_{\beta j} \, U_{\alpha j}^\ast \, e^{-i m_j^2 L/(2E)}
\right|^2 ,
\ee
and for the case of one sterile neutrino the sum 
goes from 1 to 4. For muon to tau neutrinos the probability 
is given by\footnote{Inclusion of matter effects for oscillation 
experiments with 
longer baseline will yield somewhat 
more complicated probabilities, which have however the same 
structure \cite{Ray}.} 
\bea \D \label{eq:Pst}
P(\nu_\mu \ra \nu_\tau)^{\rm St} \simeq \sin^2 2 \theta_{23} \, 
(1 - 2 \, |U_{e3}|^2) \, \sin^2 \frac{\Delta_{13}}{2} + 
\theta_{24} \, \theta_{34} \, \sin \delta_{34} \, \sin \Delta_{13} \\ \D 
- (\theta_{24}^2 + \theta_{34}^2) \, \sin^2 2 \theta_{23} \, 
\sin^2 \frac{\Delta_{13}}{2} + \ldots + 
2 \, \theta_{24}^2 \, \theta_{34}^2 \, . 
\eea
We have expanded the lengthy result in terms of the small 
parameters $|U_{e3}|$ and $\theta_{i4}$, and 
$\Delta_{31} = (m_3^2 - m_1^2) \, L/(2 E)$ corresponds to the
oscillations of the atmospheric mass-squared difference.  
The second term in 
Eq.~(\ref{eq:Pst}) is an ``interference term'' of the sterile neutrino 
mixing parameter with the leading oscillations. 
From the many higher order terms we have given here only 
the constant one, which is particularly interesting.
Namely, it introduces a constant contribution stemming purely 
from the averaged-out sterile neutrino mass-squared difference. 
This resembles the ``zero-distance'' effect in case of a non-unitary
mixing matrix \cite{old0}, which denotes a constant non-zero 
transition probability also in the limit of zero baseline. To further 
study this, consider the neutrino 
oscillation probability for a non-unitary PMNS matrix and its 
zero baseline limit:  
\be \label{eq:PNU}
P^{\rm NU}(\nu_\alpha \ra \nu_\beta) = 
\frac{
\left|\sum_j N_{\beta j} \, N_{\alpha j}^\ast \, e^{-i m_j^2 L/(2E)}
\right|^2}
{
(N N^\dagger)_{\alpha\alpha}  \, 
(N N^\dagger)_{\beta\beta}  
} 
\stackrel{L \ra 0}{=} 
\frac{\left| (N N^\dagger)_{\beta \alpha} \right|
} {
(N N^\dagger)_{\alpha\alpha}  \, 
(N N^\dagger)_{\beta\beta}  
}\, .
\ee
To be more specific, for the same 
experimental setup as before, and for the muon to tau 
channel one finds\footnote{Note that 
the sum in Eq.~(\ref{eq:PNU}) goes from 1 to 3.} 
after an expansion to second order in 
$|U_{e3}|$ and $\eta_{\mu\tau}$: 
\bea \D \label{eq:oscNU}
P(\nu_\mu \ra \nu_\tau)^{\rm NU} \simeq 
(1 - 2 \, |U_{e3}|^2) \, \sin^2 2 \theta_{23} \, \sin^2 \frac{\Delta_{31}}{2} - 
2 \, |\eta_{\mu\tau}| \, \sin 2 \phi_{\mu \tau} 
\, \sin 2 \theta_{23} \, \sin \Delta_{31} \\ \D + 
|\eta_{\mu\tau}|^2 \left( 
4 - 2 \, (3 - \cos \phi_{\mu\tau} ) \, \sin^2 \frac{\Delta_{31}}{2}  
\right) .
\eea 
We have kept only $\eta_{\mu\tau}$ in the above expression, because it is 
the leading non-unitarity parameter in $P(\nu_\mu \ra \nu_\tau)^{\rm NU}$ 
\cite{CP_Sp,GO}. 
In analogy to the expression (\ref{eq:Pst}) for sterile neutrino 
oscillations, the second term of order $|\eta_{\mu\tau}|$ 
is an interference term 
with the oscillations, while the third term of order $|\eta_{\mu\tau}|^2$ 
gives rise to a constant contribution: the zero distance effect. 
Indeed, in the limit of vanishing baseline: 
\be
P(\nu_\mu \ra \nu_\tau)^{\rm NU} 
\stackrel{L \ra 0}{\simeq}
4 \, |\eta_{\mu\tau}|^2 \, .
\ee
The structure of the two oscillation probabilities is identical, 
and governed (up to a factor 2) 
by the same element of $\zeta$ and $\eta$, respectively 
(note that $\theta_{24}^2 \, \theta_{34}^2 = 4 \, |\zeta_{\mu\tau}|^2$). 
While for the sterile neutrino case the constant term arises due to the 
averaged-out oscillation term of the large $\Delta m^2$, for non-unitarity 
the non-orthogonality of the neutrino states is the reason.\\

The same analysis can be performed for the maybe more 
motivated case \cite{boone} of two additional 
sterile neutrinos\footnote{Generalization to more sterile states 
is straightforward.}. Parameterizing 
\be
U = \tilde{R}_{15} \, \tilde{R}_{14} \, \tilde{R}_{35} \, R_{24} \, 
\tilde{R}_{25} \, \tilde{R}_{34} \, 
R_{23} \, \tilde{R}_{13} \, R_{12} \, P \, ,
\ee
one finds in the same way as above 
\[ 
\zeta \simeq -\frac 12 
\left( 
\bad 
\theta_{14}^2 + \theta_{15}^2 
& \theta_{14} \, \theta_{24} \, e^{-i \delta_{14}} + 
\theta_{15} \, \theta_{25} \, e^{i (\delta_{25} - \delta_{15})}
& \theta_{14} \, \theta_{34} \, e^{i (\delta_{34} - \delta_{14})} + 
\theta_{15} \, \theta_{35} \, e^{i (\delta_{35} - \delta_{15})} \\
\cdot & \theta_{24}^2 + \theta_{25}^2  
 &  \theta_{24} \, \theta_{34} \, e^{-i \delta_{34}} + 
\theta_{25} \, \theta_{35} \, e^{i (\delta_{35} - \delta_{25})}
\\
\cdot & \cdot  & \theta_{34}^2 + \theta_{35}^2  
\ea
\right) .
\]
The off-diagonal terms not written down explicitly are simply the 
complex conjugates of the terms from the other side of the diagonal. 
The oscillation probability reads 
\bea \D \label{eq:Pst2}
P(\nu_\mu \ra \nu_\tau)^{\rm St} \simeq \sin^2 2 \theta_{23} \, 
(1 - 2 \, |U_{e3}|^2) \, \sin^2 \frac{\Delta_{13}}{2} \\ \D + 
\left( 
\theta_{24} \, \theta_{34} \, \sin \delta_{34} 
+ \theta_{25} \, \theta_{35} \, \sin (\delta_{35} - \delta_{25})  
\right) \sin \Delta_{13} \\ \D 
- \left(
\theta_{24}^2 + \theta_{34}^2 + \theta_{25}^2 
+ \theta_{35}^2 
\right) \sin^2 2 \theta_{23} \sin^2 \frac{\Delta_{13}}{2} 
\,  ,
\eea
and is governed by the entries $\zeta_{\mu\mu}$, 
$\zeta_{\mu\tau}$ and $\zeta_{\tau\tau }$. In analogy to the 
case of one sterile neutrino, there is a constant 
``zero distance contribution'' coming from the fourth order terms, 
which read $2 \left( 
\theta_{24}^2 \, \theta_{34}^2 + 
\theta_{25}^2 \, \theta_{35}^2 \right)$.

We conclude that long baseline oscillation probabilities show similar 
properties for sterile neutrinos with mass-squared differences exceeding 
roughly 0.1 eV$^2$ and for general non-unitarity. 
Of course, the phenomenology in other sectors, even 
other oscillation experiments, is very 
different\footnote{One example in the field of neutrino 
oscillations is high energy (about TeV) long baseline 
atmospheric neutrinos 
\cite{sandhya}, in which sterile neutrinos could be detected due 
to their matter effects.}. 
In what regards mass-related observables, the eV scale states 
contribute to the cosmological sum of masses, and also to $m_\beta$ and 
\meff. For the latter two the relevant expression is 
\be
U_{e4}^2 \, m_4 = m_4 \, c_{24}^2 \, c_{34}^2 \, s_{14}^2  \, e^{i (\alpha_4 
+ 2 \delta_{14})}
\simeq m_4 \, \theta_{14}^2  \, e^{i (\alpha_4 + 2 \delta_{14})} 
\simeq 
m_4 \, e^{i (\alpha_4 + 2 \delta_{14})} \, |\zeta_{ee}|^2 \, . 
\ee
If the three active neutrinos are quasi-degenerate, $m_4$ needs to be 
larger than $m_0$ (to avoid 
too many problems with cosmology), such that there will 
be a large effect for the observables related to neutrino mass. 
The phenomenology of those scenarios is discussed e.g.~in \cite{meff_4}. 

Also in the case of two additional sterile neutrinos there is
interesting mass-related phenomenology, for which we refer to  
Ref.~\cite{meff_5}. The important quantity is 
\be
U_{e4}^2 \, m_4 + U_{e5}^2 \, m_5 \simeq 
\theta_{14}^2 \, e^{i(\alpha_4 - 2  \delta_{14})} \, m_4 + 
\theta_{15}^2 \, e^{i(\alpha_5 - 2  \delta_{15})} \, m_5 \, ,
\ee
which, unlike the 4 neutrino case, 
does not correspond directly to an element of $\zeta$, unless there is
a hierarchy in the form of $\theta_{14}^2 \ll \theta_{15}^2$ or 
$\theta_{14}^2 \gg \theta_{15}^2$.

\section{\label{sec:concl}Summary and Conclusions}
Non-unitarity of the lepton mixing matrix affects the 
observables relevant to lepton mixing and neutrino mass. 
Here we noted three possibilities related to this issue. 
We have first discussed the general case: 
for quasi-degenerate neutrinos with a 
common mass scale $m_0$, there can be corrections to the 
naive relations $\meff^{\rm max} = m_\beta = \Sigma/3$, which are
larger than the corrections stemming from the non-degeneracy 
of the masses. Second, in the context of the type I see-saw mechanism, 
it was discussed before that limits from light neutrino exchange 
in \onbb~can be translated to limits on heavy neutrino parameters. 
We showed that these limits are 
much stronger than the usual ones, which arise when one 
assumes heavy neutrino exchange in \onbb. 
The expression which makes it possible to relate the heavy neutrino 
parameters to light neutrino exchange has its origin in the 
inevitable non-unitarity of the PMNS matrix in type I see-saw
scenarios. As an example, 
the collider impact of the stronger limit on inverse neutrino-less 
double beta decay was studied, showing drastic suppression of the cross 
section.  
Finally, an analogy in neutrino oscillation probabilities between 
sterile neutrinos and non-unitarity was pointed out.

\vspace{0.3cm}
\begin{center}
{\bf Acknowledgments}
\end{center}
It is a pleasure to thank Zhi-zhong Xing for discussions and hospitality 
at the Institute of High Energy Physics of the Chinese Academy of Sciences 
in Beijing, where parts of this paper were written. Financial support  
for a research visit to Beijing from the Sino-German Center for 
Research Promotion is gratefully acknowledged. 
W.R.~was supported by the ERC under the Starting Grant 
MANITOP and by the Deutsche Forschungsgemeinschaft 
in the Transregio 27 ``Neutrinos and beyond -- weakly interacting 
particles in physics, astrophysics and cosmology''.

\end{document}